\documentclass[10pt,conference]{IEEEtran}

\usepackage{verbatim,amsmath,enumerate,amssymb}
\usepackage[square,comma,numbers]{natbib} 
\usepackage{inputenc}
\usepackage[T1]{fontenc}\usepackage{amsmath,amssymb,color,verbatim}

\def\norm#1{\left\| #1 \right\|}
\newtheorem{definition}{Definition}
\newtheorem{thm}{Theorem}

\newtheorem{lemma}{Lemma}

\usepackage{tikz}
\tikzset { domaine/.style 2 args={domain=#1:#2} }

\DeclareMathOperator*{\tr}{tr}

\providecommand{\abs}[1]{\ensuremath{\left\lvert #1 \right\rvert}}
\providecommand{\norm}[1]{\ensuremath{\left\Vert #1 \right\Vert}}

\providecommand{\floor}[1]{\ensuremath{\left\lfloor #1 \right\rfloor}}

\providecommand{\vv}[1]{\textquotedblleft #1\textquotedblright}

\newcommand{\Q}{\mathbb{Q}}
\newcommand{\Z}{\mathbb{Z}}
\newcommand{\C}{\mathbb{C}}
\newcommand{\R}{\mathbb{R}}
\newcommand{\N}{\mathbb{N}}

\newcommand{\ba}{\boldsymbol\alpha}
\renewcommand{\H}{\mathbb{H}}

\DeclareMathOperator*{\out}{out}

\providecommand{\abs}[1]{\ensuremath{\left\lvert #1 \right\rvert}}
\providecommand{\norm}[1]{\ensuremath{\left\Vert #1 \right\Vert}}

\providecommand{\vv}[1]{\textquotedblleft #1\textquotedblright}
\newcommand*{\dotleq}{\mathrel{\dot{\leq}}}
\newcommand*{\dotgeq}{\mathrel{\dot{\geq}}}

\newcommand{\D}{{\mathcal D}}

\newcommand{\mindet}[1]{\hbox{\rm det}_{min}\left( #1\right)}

\renewcommand{\IEEEQED}{\IEEEQEDopen}

\renewcommand{\IEEEQED}{\IEEEQEDopen}
\title{The DMT classification of real and quaternionic lattice codes}

\author{ 
\IEEEauthorblockN{Laura Luzzi}
\IEEEauthorblockA{ETIS -  Université Paris-Seine \\
 (Université de Cergy-Pontoise, ENSEA, CNRS) \\
Cergy-Pontoise, France \\
laura.luzzi@ensea.fr}
\and

\IEEEauthorblockN{Roope Vehkalahti}
\IEEEauthorblockA{Department of Communications and Networking\\
Aalto University\\
Helsinki, Finland\\
roope.vehkalahti@aalto.fi}

}

\begin{document}
\maketitle

\begin{abstract}
In this paper we consider space-time codes where the code-words are restricted to either real or quaternion matrices. We prove two separate diversity-multiplexing gain trade-off (DMT) upper bounds for such codes and provide a criterion for  a lattice code to achieve these upper bounds.
We also point out that lattice codes based on $\Q$-central division algebras satisfy this optimality criterion.
As a corollary this result provides a DMT classification for all $\Q$-central division algebra codes that are based on standard embeddings.

\end{abstract}

\section{Introduction}
In \cite{EKPKL} the authors proved that for every number of transmit antennas $n$ there exist a DMT optimal code in the space $M_{n}(\C)$.  These codes are derived from division algebras where the center of the division algebra is a complex quadratic field. However, this result is actually more general, and their proof revealed that as long as a $2n^2$-dimensional lattice code in  $M_n(\C)$ has the non-vanishing determinant property (NVD), it is DMT optimal. Yet, this result does not tell us  anything about space-time lattice codes that are not full dimensional in $M_n(\C)$. Such codes   naturally appear in the scenario where we have less receive than transmit antennas and try to keep the decoding complexity limited. 

One natural class of such space-time codes are the codes derived from $\Q$-central division algebras. In this paper we will measure their DMT. Unlike the case of complex quadratic center, $\Q$-central division algebras are divided into two categories with respect to their DMT performance.
This  division is  based   on the ramification of the infinite Hasse-invariant of the division algebra, which 
 decides if the lattice code corresponding to the division algebra can  be embedded into real or quaternionic space.

Our DMT classification holds for any multiplexing gain, extending previous partial results in \cite{VLL2013,ISIT2016_MIMO} which were based on the theory of Lie algebras. We note that the approach used in this paper is quite different and more general. In the spirit of \cite{EKPKL} we are  not 
just considering division algebra codes, but
all space-time codes where the code matrices are restricted to  $M_n(\R)$ (resp. $M_{n/2}(\H)$), and  provide two different upper bounds for the DMT of such codes. We then prove that if we have a  degree $n^2$-dimensional NVD lattice inside $M_n(\R)$ (resp. $M_{n/2}(\H)$) then this code achieves the respective upper bound. As the $\Q$-central division algebra codes are of this type,
we get their DMT as a corollary.

 
 

\section{Notation and preliminaries} 

\paragraph*{Notation} 
Given a matrix $X$, we denote its complex conjugate by $X^*$, its transpose by $X^T$ and its conjugate transpose by $X^{\dagger}$.\\
We use the the dotted inequality $f(\rho) \dotleq g(\rho)$ to mean $\lim_{\rho\to \infty}\frac{\log f(\rho)}{\log \rho} \leq \lim_{\rho\to \infty}\frac{\log g(\rho)}{\log \rho},$ and similarly for equality.

\subsection{Subspaces and lattices}\label{basic}
 In this paper we will consider space-time codes  that are subsets of  certain subspaces of the  $2n^2$-dimensional real vector space $M_n(\C)$.  The first such subspace consists of all the  real matrices  inside $M_n(\C)$ and we denote it with $M_n(\R)$. The other subspace of interest consists of quaternionic matrices.

Let us assume that $2\mid n$. We denote  with  $M_{n/2}(\H)$ the set of quaternionic matrices 
$$
\begin{pmatrix}
A &   -B^* \\
B&  A^*                                               
\end{pmatrix}
\in M_{n}(\C),
$$
where $*$ refers to complex conjugation and $A$ and $B$ are complex matrices in $M_{n/2}(\C)$. 
Note that  quaternionic matrices form a  $n^2$-dimensional subspace in $M_n(\C)$.

The space-time codes we consider in this work are based on additive groups in $M_{n}(\C)$.
\begin{definition}
A {\em matrix lattice} $L \subseteq M_{n}(\C)$ has the form
$$
L=\Z B_1\oplus \Z B_2\oplus \cdots \oplus \Z B_k,
$$
where the matrices $B_1,\dots, B_k$ are linearly independent over $\R$, i.e., form a lattice basis, and $k$ is
called  the \emph{dimension} of the lattice.
\end{definition}
We immediately see that if we have a lattice inside the space $M_n(\R)$ or $M_{n/2}(\H)$ the maximal  dimension it can have is $n^2$.

\begin{definition}\label{def:NVD}
If the \emph{minimum determinant} of the lattice $L \subseteq M_{n\times n}(\C)$ is non-zero, i.e. satisfies
\[
\mindet{L}:=\inf_{{\bf 0} \neq X \in L}\abs{\det (X)} > 0, 
\]
we say that the lattice satisfies the \emph{non-vanishing determinant} (NVD) property.
\end{definition}

Building high dimensional NVD lattices is a highly non-trivial task. A natural source of such lattices are division algebras. 
Let  $\mathcal{D}$ be a degree $n$   $\Q$-central division algebra.
We say that the algebra $\D$ is \emph{ramified at the infinite place} if 
$
\D\otimes_{\Q}\R\simeq M_{n/2}(\mathbb{H})$.
If it is not, then
$
\D\otimes_{\Q}\R\simeq M_{n}(\R).
$

Let $\Lambda$ be an \emph{order} in $\mathcal{D}$.
\begin{lemma} \cite[Lemma 9.10]{VLL2013}\label{embeddings}
If the infinite prime is ramified in the algebra $\D$, then there exists an embedding
$$\psi: \mathcal{D} \to M_{n/2}(\mathbb{H})$$
 such that $\psi(\Lambda)$ is a $n^2$ dimensional NVD lattice.
If $\D$ is not ramified at the infinite place, then there exists an embedding
$$\psi: \mathcal{D} \to M_{n}(\R)$$
such that $\psi(\Lambda)$ is a $n^2$ dimensional NVD lattice.
\end{lemma}

\subsection{Channel model}
We consider a MIMO system with $n$ transmit and $m$ receive antennas, and minimal delay $T=n$. 
The received signal is 
\begin{equation} \label{channel}
Y_c=\sqrt{\frac{\rho}{n}} H_c \bar{X} + W_c,
\end{equation}
where $\bar{X} \in M_n(\C)$ is the transmitted codeword, $H_c \in M_{m,n}(\C)$ and $W_c \in M_{m,n}(\C)$ are the channel and noise matrices with i.i.d. circularly symmetric complex Gaussian entries $h_{ij}, w_{ij} \sim \mathcal{N}_{\C}(0,1)$, and $\rho$ is the signal-to-noise ratio (SNR). The set of transmitted codewords $\mathcal{C}$ satisfies the average power constraint
\begin{equation} \label{power_constraint}
\frac{1}{\abs{\mathcal{C}}} \frac{1}{n^2} \sum_{X \in \mathcal{C}} \norm{X}^2 \leq 1.
\end{equation} 
We suppose that perfect channel state information is available at the receiver but not at the transmitter, and that maximum likelihood decoding is performed. 

In the DMT setting \cite{ZT}, we consider codes $\mathcal{C}(\rho)$ whose size grows with the SNR, and define the multiplexing gain as 
$$r=\lim_{\rho \to \infty} \frac{1}{n}\frac{\log \abs{\mathcal{C}}}{\log \rho},$$
and the diversity gain as 
$$d(r)=-\lim_{\rho \to \infty}\frac{\log P_e}{\log \rho},$$
where $P_e$ is the average error probability.
\paragraph*{Spherically shaped lattice codes} 
Let now $L$ be  a lattice in $M_n(\C)$. Given $M$, consider the subset of elements whose Frobenius norm is bounded by $M$:
$$L(M)=\{ X \in L \;:\; \norm{X} \leq M\}.$$
Let $k \leq 2n^2$ be the dimension of $L$ as a $\Z$-module. As in \cite{VLL2013}, we choose $M=\rho^{\frac{rn}{k}}$ and consider codes of the form 
\begin{equation*} 
\mathcal{C}(\rho)=M^{-1} L(M)=\rho^{-\frac{rn}{k}} L(\rho^{\frac{rn}{k}}),
\end{equation*}
which satisfy the power constraint (\ref{power_constraint}). The multiplexing gain of this code is $r$.


\section{Real lattice codes}
In this section, we focus on the special case where $\mathcal{C}(\rho) \subset M_n(\R)$, i.e. the code is a set of real matrices. 
\subsection{Equivalent real channel} First, we show that the channel model (\ref{channel}) is equivalent to a real channel with $n$ transmit and $2m$ receive antennas.\\
We can write $H_c=H_r + i H_i$, $W_c=W_r + i W_i$, where $H_r,H_i,W_r,W_i$ have i.i.d. real Gaussian entries with variance $1/2$. If $Y_c=Y_r + i Y_i$, with $Y_r, Y_i \in M_{m \times n}(\R)$, we can write an equivalent real system with $2m$ receive antennas:
\begin{equation} \label{real_channel}
Y=\begin{pmatrix} Y_r \\ Y_i \end{pmatrix} = 
\sqrt{\frac{\rho}{n}} \begin{pmatrix} H_r \\ H_i \end{pmatrix} \bar{X} + \begin{pmatrix} W_r \\ W_i \end{pmatrix} =\sqrt{\frac{\rho}{n}} H\bar{X} + W,
\end{equation}
where $H \in M_{2m \times n}(\R)$, $W \in M_{2m \times n}(\R)$ have real i.i.d. Gaussian entries with variance $1/2$.

\subsection{General DMT upper bound for real codes} Using the equivalent real channel in the previous section, we can now establish a general upper bound for the DMT of real codes.

\begin{thm} \label{theorem_real_upper}
Suppose that $\forall \rho$, $\mathcal{C}(\rho) \subset M_n(\R)$. Then the DMT of the code $\mathcal{C}$ is upper bounded by the function $d_1(r)$ connecting the points $(r,[(m-r)(n-2r)]^+)$ where $2r \in \Z$.
\end{thm}

\begin{IEEEproof} This part of the proof closely follows \cite{ZT}. Given a rate $R=r \log \rho$, consider the outage probability \cite{Telatar}
\begin{equation} \label{P_out}
P_{\out}(R)=\inf_{Q \succ 0, \;\tr(Q) \leq n} \mathbb{P}\left\{ \Psi(Q,H) \leq R\right\},
\end{equation}
where $\Psi(Q,H)$ is the maximum mutual information per channel use of the real MIMO channel (\ref{real_channel}) with fixed $H$ and real input with fixed covariance matrix $Q$.\footnote{Unlike \cite{Telatar} and \cite{ZT}, we don't use a strict inequality in the definition (\ref{P_out}), but our definition is equivalent since the set of $H$ such that $\Psi(Q,H)=R$ has measure zero.} 
  Following a similar reasoning as in \cite[Section 3.2]{Telatar}, it is not hard to see that 
$$\Psi(Q,H)=\frac{1}{2} \log \det (I + \frac{\rho}{n} H Q H^T).$$ 
As in \cite[Section III.B]{ZT}, since $\log \det$ is increasing on the cone of positive definite symmetric matrices, for all $Q$ such that $\tr(Q)\leq n$ we have $\frac{Q}{n}\preceq I$ and
$$P_{\out}(R) \geq \mathbb{P} \left\{\frac{1}{2} \log \det (I + \rho HH^T)  \leq R\right\}.$$
Note that $\det(I+\rho HH^T)=\det(I+\rho H^T H)$. Let $l=\min(2m,n)$, and $\Delta=\abs{n-2m}$. Let $\lambda_1 \geq \lambda_2 \geq \cdots \geq \lambda_l > 0$ be the nonzero eigenvalues of $H^T H$. The joint probability distribution of  $\boldsymbol\lambda=(\lambda_1,\ldots,\lambda_l)$ is given by \cite{Edelman}\footnote{We have slightly modified the expression to be consistent with our notation. In \cite{Edelman}, the author considers a matrix $A^TA$ where each element of $A$ is $\mathcal{N}(0,1)$.}:
\begin{equation} \label{p_lambda_real}
p(\boldsymbol\lambda)=Ke^{-\sum\limits_{i=1}^l \lambda_i} \prod_{i=1}^l \lambda_i^{\frac{\Delta-1}{2}}\prod_{i<j}(\lambda_i-\lambda_j)
\end{equation}
for some constant $K$.
 Consider the change of variables $\lambda_i=\rho^{-\alpha_i} \;\forall i$.  The corresponding distribution for $\boldsymbol\alpha=(\alpha_1,\ldots,\alpha_l)$ in the set  $\mathcal{A}=\{\ba\;:\; \alpha_1 \leq \cdots \leq \alpha_l\}$ is
\begin{equation} \label{p_alpha_real}
\!p(\boldsymbol\alpha)\!=\!K(\log \rho)^l e^{-\!\!\sum\limits_{i=1}^l \! \rho^{-\alpha_i}}\!\rho^{-\!\!\sum\limits_{i=1}^l\!\alpha_i \left(\frac{\Delta+1}{2}\right)}\!\prod_{i<j}\!\left(\rho^{-\alpha_i}\!\!-\!\rho^{-\alpha_j}\!\right) 
\end{equation}
Then we have
{\allowdisplaybreaks
\begin{align*}
&P_{\out}(R) \doteq \mathbb{P}\left\{ \prod_{i=1}^l (1+ \rho \lambda_i)  \leq \rho^{2r}\right\}\\
&=\mathbb{P}\left\{ \prod_{i=1}^l(1+\rho^{1-\alpha_i}) \leq \rho^{2r}\right\}.
\end{align*}
}%
To simplify notation, we take $s=2r$. Note that $1+\rho^{1-\alpha_i} \leq 2 \rho^{(1-\alpha_i)^+} \doteq \rho^{(1-\alpha_i)^+}$, therefore
{\allowdisplaybreaks
\begin{align*}
& P_{\out}(R) \dotgeq \mathbb{P}\left\{ \prod_{i=1}^l \rho^{(1-\alpha_i)^{+}} \leq  \rho^s\right\}
\geq \mathbb{P}(\mathcal{A}_0),
\end{align*}
}%
where
{
\begin{align} \label{A_0} 
&\mathcal{A}_0=
\left\{\ba \in \mathcal{A}:\;\alpha_i \geq 0 \;\forall i=1,\ldots,l,\; \sum_{i=1}^l (1-\alpha_i)^+  \leq s\right\} \notag\\
&=\!\left\{\!\ba \in \mathcal{A}\!:\;\!\alpha_j\geq 0,\;\sum_{i=1}^j (1-\alpha_i) \leq s \; \forall j=1,\ldots,l\right\}.
\end{align}
}%
In fact, given $\ba \in \mathcal{A}$, let $t=t(\ba)$ be such that $\alpha_{t+1} \geq 1 \geq \alpha_t$. Then $\forall j=1,\ldots,l$,
$\sum_{i=1}^j (1-\alpha_i) \leq \sum_{i=1}^t (1-\alpha_i)=\sum_{i=1}^l (1-\alpha_i)^+.$\\
Consider $S_{\delta}=\{\ba \in \mathcal{A}:\; \abs{\alpha_i-\alpha_j}> \delta \; \forall i \neq j\}$. Then 
{\allowdisplaybreaks
\begin{align*}
&P_{\out}(R) \dotgeq \int_{\mathcal{A}_0} e^{-\sum\limits_{i=1}^l \rho^{-\alpha_i}} \rho^{- \sum\limits_{i=1}^l\frac{(\Delta+1)\alpha_i}{2}} \prod_{i<j} (\rho^{-\alpha_i}-\rho^{-\alpha_j}) d \ba\\
&\geq \int_{\mathcal{A}_0 \cap S_{\delta}}e^{-\sum\limits_{i=1}^l \rho^{-\alpha_i}} \rho^{-\sum\limits_{i=1}^l \frac{(\Delta+1)\alpha_i}{2}} \prod_{i<j} (\rho^{-\alpha_i}-\rho^{-\alpha_j}) d \ba \\
&\geq \frac{(1-\rho^{-\delta})^l}{e^l}  \int_{\mathcal{A}_0 \cap S_{\delta}}  \rho^{-\sum\limits_{i=1}^l\alpha_i N_i} d\ba \doteq  \int_{\mathcal{A}_0 \cap S_{\delta}} \rho^{-\sum\limits_{i=1}^l\alpha_i N_i} d\ba,
\end{align*} 
}%
where $N_i=\frac{\Delta+2l-2i+1}{2}$. The previous inequality follows from the fact that $\rho^{-\alpha_i}-\rho^{-\alpha_j}>\rho^{-\alpha_i}(1-\rho^{-\delta})$ for $\ba \in S_{\delta}$, and $e^{-\sum\limits_{i=1}^l \rho^{-\alpha_i}}\geq \frac{1}{e}$ if $\alpha_i \geq 0$. (Note that for a fixed $i$, there are $l-i$ possible values for $j$ such that $i<j$.)

\begin{lemma} \label{inf_lemma}
Let $f(\ba)=\sum\limits_{i=1}^l (q+l+1-2i)\alpha_i$. 
Then 
$$\inf\limits_{\ba \in \mathcal{A}_0}  f(\ba)=(-q-l+2\floor{s}+1)s+ql-\floor{s}(\floor{s}+1)=f(\boldsymbol\alpha^*),$$
where $\alpha_1^*=\ldots=\alpha_{k-1}^*=0$, $\alpha_k^*=k-s$, $\alpha_{k+1}^*=\ldots=\alpha_l^*=1$.
\end{lemma}

The proof of Lemma \ref{inf_lemma} can be found in Appendix \ref{proof_inf_lemma}.\\
Using Lemma \ref{inf_lemma} with $q=\Delta+l$, $s=2r$, we find that $\inf_{\ba \in \mathcal{A}_0} \sum_{i=1}^l N_i \alpha_i=\inf_{\ba \in \mathcal{A}_0} \frac{f(\ba)}{2}$ is equal to
{\allowdisplaybreaks
\begin{align*}
&\frac{1}{2}\left[(-\Delta-2l+2\floor{2r}+1)2r+(\Delta+l)l-\floor{2r}(\floor{2r}+1)\right]\\
&=(-2m-n+2\floor{2r}+1)r+mn-\frac{\floor{2r}(\floor{2r}+1)}{2}.
\end{align*}
}%
This is the piecewise function $d_1(r)$ connecting the points $(r,[(m-r)(n-2r)]^+)$ where $2r \in \Z$.\\
Using the Laplace principle, $\forall \delta>0$ we have
$$\lim_{\rho \to \infty} -\frac{\log P_{\out}(R)}{\log\rho}\geq \inf_{\mathcal{A}_0 \cap S_{\delta}} \frac{f(\boldsymbol\alpha)}{2}.$$
Note that $\forall \delta$, the point
$\ba_{\delta}$ such that $\alpha_{\delta,i}=\alpha_i^*+\frac{\delta i}{l}$ is in $\mathcal{A}_0 \cap S_{\frac{\delta}{l}}$ and when $\delta \to 0$, $\ba_{\delta} \to \ba^*$. By continuity of $f$,
\begin{align}
\lim_{\delta \to 0} \inf_{\mathcal{A}_0 \cap S_{\delta}} \frac{f(\ba)}{2}=\frac{f(\ba^*)}{2}=d_1(r). \tag*{\IEEEQED}
\end{align}
\let\IEEEQED \relax%
\end{IEEEproof}

\subsection{DMT of real lattice codes with NVD}
In this section, we show that real spherically shaped lattice codes with the NVD property achieve the DMT upper bound of Theorem \ref{theorem_real_upper}. This result extends Proposition 4.2 in \cite{ISIT2016_MIMO}.
\begin{thm} \label{theorem_real_lower}
Let $L$ be an $n^2$-dimensional lattice in $M_n(\R)$, and consider the spherically shaped code $\mathcal{C}(\rho)=\rho^{-\frac{r}{n}} L(\rho^{\frac{r}{n}})$.\\
If $L$ has the NVD property, then the DMT of the code $\mathcal{C}(\rho)$ is 
the function $d_1(r)$ connecting the points $(r,[(m-r)(n-2r)]^+)$ where $2r \in \Z$.
\end{thm}

\begin{IEEEproof}
Since the upper bound has already been established in Theorem \ref{theorem_real_upper}, we only need to prove that the DMT is lower bounded by $d_1(r)$. The following section follows very closely the proof in \cite{EKPKL}, and thus some details are omitted. To simplify notation, we assume that $\mindet{L}=1$.\\
We consider the sphere bound for the error probability for the equivalent real channel (\ref{real_channel}): for a fixed channel realization $H$,
$$P_e(H) \leq \mathbb{P}\left\{ \norm{W}^2 > d_H^2/4\right\}$$
where $d_H^2$ is the 
squared 
minimum distance in the received constellation: 
\begin{align*}
&d_H^2\doteq\rho \min_{\bar{X},\bar{X}' \in \mathcal{C}(\rho),\;\bar{X} \neq \bar{X}'} \norm{H(\bar{X}-\bar{X}')}^2\\
&=\rho^{1-\frac{2r}{n}} \min_{X,X' \in L(\rho^{\frac{r}{n}}),\;X \neq X'} \norm{H(X-X')}^2.  
\end{align*}
We denote $\Delta X=X-X'$. Let $l=\min(2m,n)$, and $\Delta=\abs{n-2m}$. Let $\lambda_1 \geq \lambda_2 \geq \cdots \geq \lambda_l >0$ be the non-zero eigenvalues of $H^T H$, and $0 \leq \mu_1 \leq \cdots \leq \mu_n$ 
the eigenvalues of $\Delta X \Delta X^T$. Using the mismatched eigenvalue bound and the arithmetic-geometric inequality as in \cite{EKPKL}, 
for all $k=1, \ldots, l$
{\allowdisplaybreaks
\begin{align*}
&d_H^2\doteq\rho^{1-\frac{2r}{n}}\min_{X,X' \in L(\rho^{\frac{r}{n}}),\;X \neq X'} \tr(H\Delta X \Delta X^T H^T)\\
&\geq \rho^{1-\frac{2r}{n}} \sum_{i=1}^l \mu_i \lambda_i \geq k \rho^{1-\frac{2r}{n}}  \left(\prod_{i=1}^k \lambda_i\right)^{\frac{1}{k}} \left(\prod_{i=1}^k \mu_i\right)^{\frac{1}{k}}.
\end{align*} 
}
For all $i= 1, \ldots, n$, $\mu_i \leq \norm{\Delta X}^2 \leq 
4 \rho^{\frac{2r}{n}}$, and 
$$ \prod_{i=1}^n \mu_i =\det(\Delta X \Delta X^T) \geq 1$$
due to the NVD property. Consequently, for all $k=1,\ldots,l$
$$\prod_{i=1}^k \mu_i =\frac{\det(\Delta X \Delta X^T)}{\prod_{j=k+1}^n \mu_j} \geq \frac{1}{\rho^{\frac{2r(n-k)}{n}}}.$$ 
With the change of variables $\lambda_i=\rho^{-\alpha_i}$ $\forall i=1,\ldots,l$, we can write 
{\allowdisplaybreaks
\begin{align*}
&d_H^2 \dotgeq  \rho^{1-\frac{2r}{n}} \rho^{-\frac{1}{k} \sum\limits_{i=1}^k \alpha_i} \frac{1}{\rho^{\frac{2r(n-k)}{n}}}= \rho^{-\frac{1}{k}\left(\sum\limits_{i=1}^k \alpha_i +2r -k \right)}\\
&= \rho^{\delta_k(\boldsymbol\alpha,2r)} \quad \forall k=1,\ldots,l,
\end{align*}
}
where we have set $\boldsymbol\alpha=(\alpha_1,\ldots,\alpha_l)$ and 
\begin{equation} \label{delta}
\delta_k(\boldsymbol\alpha,s)=-\frac{1}{k}\left(\sum\limits_{i=1}^k \alpha_i +s -k \right).
\end{equation}
To simplify the notation, we will take $s=2r$.\\
Since $2 \norm{W}^2$ is a $\chi^2(2mn)$ random variable, we have
$$\mathbb{P}\left\{\norm{W}^2 >d\right\}=\sum_{j=0}^{mn-1}e^{-d} \frac{d^j}{j!}.$$
Let $p(\ba)$ be the distribution of $\ba$ in (\ref{p_alpha_real}). Note that for $i<j$, $\rho^{-\alpha_i} \geq \rho^{-\alpha_j}$ and for a fixed $i$, there are $l-i$ possible values for $j$. Consequently
\begin{equation} \label{p_prime}
p(\boldsymbol\alpha) \leq p'(\boldsymbol\alpha)=K e^{-\sum\limits_{i=1}^l \rho^{-\alpha_i}} \rho^{-\sum\limits_{i=1}^l\alpha_i N_i}(\log \rho)^l
\end{equation}
where $N_i=\frac{\Delta+2l-2i+1}{2}$. By averaging over the channel, the error probability is bounded by
$$P_e =\int P_e(\boldsymbol\alpha) p(\boldsymbol\alpha) d\boldsymbol\alpha \leq \int \mathbb{P}\left\{ \norm{W}^2 > \frac{\rho^{\delta_k(\boldsymbol\alpha,s)}}{4}\right\} p(\boldsymbol\alpha) d\boldsymbol\alpha.$$
Finally, we get $\forall k=1,\ldots,l$,
\begin{equation} \label{integral_A}
P_e \leq  \int_{\mathcal{A}} p'(\boldsymbol\alpha) \Phi(d_H^2)d\boldsymbol\alpha \leq\int_{\mathcal{A}} p'(\boldsymbol\alpha) \Phi(\rho^{\delta_k(\ba,s)})d\boldsymbol\alpha
\end{equation}
where $\mathcal{A}=\{\ba\;:\; \alpha_1 \leq \cdots \leq \alpha_l\}$, and
\begin{equation} \label{Phi}
\Phi(d)= \mathbb{P}\left\{ \norm{W}^2 > \frac{d}{4}\right\}= e^{-\frac{d}{4}} \sum_{j=0}^{2mn-1} \left(\frac{d}{4}\right)^j\frac{1}{j!}.
\end{equation}

The following Lemma 
is proven in Appendix \ref{proof_laplace_lemma}:

\begin{lemma} \label{laplace_lemma}
{\allowdisplaybreaks
\begin{align*}
&\min_{k=1,\ldots,l}\left(-\lim_{\rho \to \infty} \frac{1}{\log \rho} \log \int_{\mathcal{A}} p'(\ba) \Phi(\rho^{\delta_k(\ba,s)}) d\ba\right) \\
&\geq \inf_{\ba \in \mathcal{A}_0} \sum_{i=1}^l N_i \alpha_i,
\end{align*}
}
where $\mathcal{A}_0$ is defined in (\ref{A_0}). 
\end{lemma}

The proof of the Theorem is concluded using Lemma \ref{inf_lemma} with $q = \Delta+l$, $s = 2r$.
\end{IEEEproof}
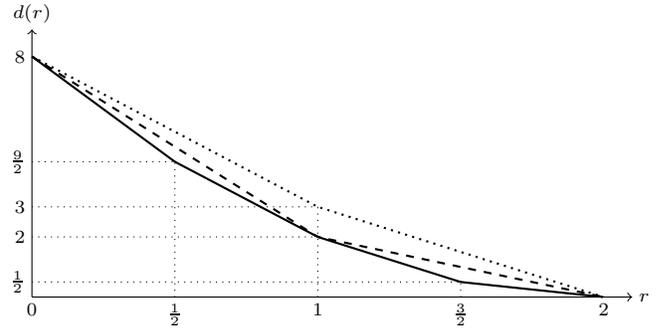
\begin{figure}[tb]
\begin{scriptsize}
\begin{tikzpicture}[xscale=3.8,yscale=0.4]
\draw[->] (0,0) -- (2.1,0);
\draw (2.1,0) node[right] {$r$};
\draw [->] (0,0) -- (0,8.9);
\draw (0,0) node[below] {$0$};
\draw (0.5,0) node[below] {$\frac{1}{2}$};
\draw (1,0) node[below] {$1$};
\draw (0,0.5) node[left] {$\frac{1}{2}$};
\draw (0,2) node[left] {$2$};
\draw (0,3) node[left] {$3$};
\draw (0,4.5) node[left] {$\frac{9}{2}$};
\draw (1.5,0) node[below] {$\frac{3}{2}$};
\draw (2,0) node[below] {$2$};
\draw (0,8) node[left] {$8$};
\draw (0,8.9) node[above] {$d(r)$};
\draw [dotted] (0.5,0) -- (0.5,4.5);
\draw [dotted] (1,0) -- (1,3);
\draw [dotted] (0,0.5) -- (1.5,0.5) ;
\draw [dotted] (1.5,0) -- (1.5,0.5);
\draw [dotted] (0,2) -- (1,2);
\draw [dotted] (0,3) -- (1,3);
\draw [dotted] (0,4.5) -- (0.5,4.5);
\draw [thick, samples=200, domain=0:2] plot(\x,{(-7+2*floor(2*\x))*\x+8-floor(2*\x)*(floor(2*\x)+1)/2});
\draw [dashed, thick, samples=200, domain=0:2] plot(\x,{2*((-1)*(3-2*floor(\x))*\x+4-floor(\x)*(floor(\x)+1))});
\draw [dotted, thick, samples=200, domain=0:2] plot(\x,{(-5+2*floor(\x))*\x+8-floor(\x)*(floor(\x)+1)});
\end{tikzpicture}
\end{scriptsize}
\caption{DMT upper bounds for real (solid) and quaternion (dashed) codes for $n=4$ and $m=2$. The dotted lines correspond to the 
optimal DMT.} 
\end{figure}

\section{Quaternion lattice codes}
Suppose that $n=2p$ is even. We consider again the channel 
\begin{equation} \label{channel_2}
Y_c=\sqrt{\frac{\rho}{n}} H_c \bar{X} + W_c,
\end{equation}
and we suppose that the codewords $\bar{X}$ are of the form
$$\bar{X}=\begin{pmatrix} A & -B^* \\ B & A^* \end{pmatrix} \in M_{2p}(\C),$$
where $A,B \in M_p(\C)$.

\subsection{Equivalent quaternion channel}
First, we derive an equivalent model where the channel has quaternionic form. We can write 
$$Y_c=\begin{pmatrix} Y_1 & Y_2\end{pmatrix}, \quad H_c=\begin{pmatrix} H_1 & H_2\end{pmatrix}, \quad W_c=\begin{pmatrix} W_1 & W_2\end{pmatrix},$$
where $Y_1, Y_2, H_1, H_2, W_1, W_2 \in M_{m \times p}(\C)$. Then 
$$Y_1\!=\!\sqrt{\frac{\rho}{n}}(H_1 A + H_2 B) + W_1, \; Y_2\!=\!\sqrt{\frac{\rho}{n}}(-H_1 B^* + H_2 A^*)+W_2,$$
and we have the equivalent \vv{quaternionic channel}:
\begin{small}
$$\underbrace{\begin{pmatrix} Y_1 & Y_2 \\ -Y_2^* & Y_1^* \end{pmatrix}}_{\text{\normalsize{$Y$}}} =\sqrt{\rho} \underbrace{\begin{pmatrix} H_1 & H_2 \\ -H_2^* & H_1^* \end{pmatrix}}_{\text{\normalsize{$H$}}} \underbrace{\begin{pmatrix} A & -B^* \\ B & A^* \end{pmatrix}}_{\text{\normalsize{$X$}}} + \underbrace{\begin{pmatrix} W_1 & W_2 \\ -W_2^* & W_1^* \end{pmatrix}}_{\text{\normalsize{$W$}}}$$
\end{small}

\subsection{General DMT upper bound for quaternion codes}

\begin{thm} \label{theorem_quaternion_upper}
Suppose that $\forall \rho$, $\mathcal{C}(\rho) \subset M_{n/2}(\H)$. Then the DMT of the code $\mathcal{C}$ is upper bounded by the function $d_2(r)$ connecting the points $(r,[(m-r)(n-2r)]^+)$ for $r \in \Z$.
\end{thm}

\begin{IEEEproof} The quaternionic channel can be written in the complex MIMO channel form
\begin{equation} \label{quaternion_channel_2}
\begin{pmatrix} Y_1 \\ -Y_2^* \end{pmatrix} =\sqrt{\frac{\rho}{n}} \begin{pmatrix} H_1 & H_2 \\ -H_2^* & H_1^* \end{pmatrix} \begin{pmatrix} A \\ B \end{pmatrix} + \begin{pmatrix} W_1 \\ -W_2^* \end{pmatrix}
\end{equation}
If $r$ is the multiplexing gain of the original system (\ref{channel_2}), then the multiplexing gain of this channel is $2r$, since the same number of symbols is transmitted using half the frame length.\\
Consider the eigenvalues  $\lambda_1=\lambda_1' \geq \lambda_2=\lambda_2' \geq \cdots \geq \lambda_p=\lambda_p' \geq 0$ of $H^{\dagger}H$. Let $l=\min(m,p)$ the number of pairs of nonzero  eigenvalues, and $\Delta=\abs{p-m}$.
For fixed $H$, the capacity of this channel is \cite{Telatar}
$$C(H) \doteq \log \det (I+ \rho H^{\dagger} H)=2 \sum_{i=1}^p \log (1+\rho \lambda_i ).$$
The joint eigenvalue density $p(\boldsymbol\lambda)=p(\lambda_1,\ldots,\lambda_l)$ of a quaternion Wishart matrix is \cite{Edelman_Rao}\footnote{The quaternion case corresponds to taking $\beta=4$ in \cite[equation (4.5)]{Edelman_Rao}. Note that we modify the distribution to take into account the fact that each entry of $H$ has variance $1/2$ per real dimension.}
$$p(\lambda_1,\ldots,\lambda_p)=K \prod_{i<j} (\lambda_i -\lambda_j)^4 \prod_{i=1}^l \lambda_i^{2\Delta+1}e^{-\sum\limits_{i=1}^l \lambda_i}$$  
for some constant $K$. Considering the change of variables $\lambda_i=\rho^{-\alpha_i}$ $\forall i=1,\ldots,l$, the distribution of $\ba=(\alpha_1,\ldots,\alpha_l)$ is
$$p(\ba)\!\!=\!\!K(\log \rho)^le^{-\sum\limits_{i=1}^l \rho^{-\alpha_i}}\! \rho^{-2\sum\limits_{i=1}^l \alpha_i (\Delta+1)}\!\prod_{i<j}\!\left(\rho^{-\alpha_i}\!\!-\!\!\rho^{-\alpha_j}\right)^4$$%
The output probability for rate $R=r \log \rho$ is given by
{\allowdisplaybreaks
\begin{align*}
&P_{\out}(R)\doteq \mathbb{P}\left\{2 \sum_{i=1}^l \log(1+\rho \lambda_i)<2r \log \rho\right\}\\
&=\!\mathbb{P}\left\{\prod_{i=1}^l (1\!+\!\rho^{1-\alpha_i})\!<\!\rho^{r}\!\right\}\!\doteq\! \mathbb{P}\left\{\prod_{i=1}^l \rho^{(1-\alpha_i)^+}\!\!<\!\rho^{r}\!\right\} \!\geq\! \mathbb{P}(\mathcal{A}_0)
\end{align*}
}%
where $\mathcal{A}_0=\left\{\ba: 0 \leq \alpha_1 \leq \ldots \leq \alpha_l,\; \sum_{i=1}^l (1-\alpha_i)^+ <r\right\}$.
Given $\delta >0$, define $S_{\delta}=\{\ba:\; \abs{\alpha_i-\alpha_j}> \delta \; \forall i \neq j\}$. Then
{\allowdisplaybreaks
\begin{align*}
&P_{\out}(R)\\ 
& \dotgeq \!\!\int_{\mathcal{A}_0 \cap S_{\delta}}\! e^{-\sum\limits_{i=1}^l \rho^{-\alpha_i}} \!\rho^{-2\sum\limits_{i=1}^l\alpha_i(\Delta +1)} \prod_{i<j} (\rho^{-\alpha_i} -\rho^{-\alpha_j})^4 d\ba\\
&\geq \frac{(1-\rho^{-\delta})^l}{e^{l}} \int_{\mathcal{A}_0 \cap S_{\delta}} \rho^{-2\sum\limits_{i=1}^l N_i \alpha_i} d\ba
\end{align*}}%
where $N_i=2(\Delta+2l-2i+1)$.
Let $f(\ba)=\sum_{i=1}^l \alpha_i N_i$. Using the Laplace principle, $\lim_{\rho \to \infty} -\frac{\log P_{\out}(R)}{\log\rho}\geq 2\inf_{\mathcal{A}_0 \cap S_{\delta}} f(\alpha)$ \; $\forall \delta>0.$
Using Lemma \ref{inf_lemma} with $s=r$, $q=\Delta+l$, we find that $2\inf_{\ba \in \mathcal{A}_0} f(\ba)=2f(\ba^*)$ is the piecewise linear function $d_2(r)$ connecting the points 
$(r, \left[2(p-r)(m-r)\right]^+)=(r,\left[(n-2r)(m-r)\right]^+)$ for $r \in \Z$. Note that $\forall \delta$, the point
$\ba_{\delta}$ such that $\ba_{\delta,i}=\alpha_i^*+\frac{\delta i}{l}$ is in $\mathcal{A}_0 \cap S_{\frac{\delta}{l}}$ and when $\delta \to 0$, $\ba_{\delta} \to \ba^*$. By continuity of $f$, $2\lim_{\delta \to 0} \inf_{\mathcal{A}_0 \cap S_{\delta}} f(\ba)=2f(\ba^*)=d_2(r)$.
\end{IEEEproof}

\subsection{DMT of quaternionic lattice codes with NVD}
 
We now show that quaternionic lattice codes with NVD achieve the upper bound of Theorem \ref{theorem_quaternion_upper}.  This result extends Proposition 4.3 in \cite{ISIT2016_MIMO}. 
 
\begin{thm} \label{theorem_quaternion_lower}
Let $L$ be an $n^2$-dimensional lattice in $M_{n/2}(\H)$, and consider the spherically shaped code $\mathcal{C}(\rho)=\rho^{-\frac{r}{n}} L(\rho^{\frac{r}{n}})$. If $L$ has the NVD property, then the DMT of the code $\mathcal{C}(\rho)$ is 
the piecewise linear function $d_2(r)$ connecting the points $(r,[(m-r)(n-2r)]^+)$ for $r \in \Z$.
\end{thm}

\begin{IEEEproof}
To simplify notation, assume $\mindet{L}=1$. 
For a fixed 
realization $H$, $P_e(H) \leq \mathbb{P}\left\{ \norm{W}^2 > d_H^2/4\right\}$,
where 
{\allowdisplaybreaks
\begin{align*}
&d_H^2 \doteq\rho^{1-\frac{2r}{n}} \min_{X,X' \in L(\rho^{\frac{r}{n}}),\;X \neq X'} \norm{H(X-X')}^2.
\end{align*}
}%
Let $\Delta X=X-X'$. 
We denote by $\lambda_1=\lambda_1' \geq \lambda_2=\lambda_2' \geq \cdots \geq \lambda_p=\lambda_p' \geq 0$  the eigenvalues of $H^{\dagger} H$, and by $0 \leq \mu_1=\mu_1' \leq \cdots \leq \mu_p=\mu_p'$ the eigenvalues of $\Delta X \Delta X^{\dagger}$. Both sets of eigenvalues have multiplicity $2$ since $H$ and $X$ are quaternion matrices. Again we set $l=\min(m,p)$ and $\Delta=\abs{p-m}$.\\
Using the mismatched eigenvalue bound and the arithmetic-geometric inequality as in \cite{EKPKL}, we find that for all $k=1,\ldots,l$,
{\allowdisplaybreaks
\begin{align*}
&d_H^2\doteq\rho^{1-\frac{2r}{n}} \min_{X,X' \in \mathcal{C}(\rho),\;X \neq X'} \tr(H\Delta X \Delta X^{\dagger} H^{\dagger}) \\
&\geq \rho^{1-\frac{2r}{n}} \sum_{i=1}^l (2 \mu_i \lambda_i) \geq 2k \rho^{1-\frac{2r}{n}} \left(\prod_{i=1}^k \lambda_i\right)^{\frac{1}{k}} \left(\prod_{i=1}^k \mu_i\right)^{\frac{1}{k}}. 
\end{align*}
}
As before, for all $i=1,\ldots,n$, $\mu_i \leq \norm{\Delta X}^2 \leq 4 \rho^{\frac{2r}{n}}$, and 
$ \prod_{i=1}^n \mu_i =\det(\Delta X \Delta X^{\dagger})^{\frac{1}{2}} \geq 1$
using the NVD property of the code. Consequently, for all $k=1,\ldots,l$
$$\prod_{i=1}^k \mu_i =\frac{\det(\Delta X \Delta X^{\dagger})^{\frac{1}{2}}}{\prod_{j=k+1}^n \mu_j} \geq \frac{1}{\rho^{\frac{2r(p-k)}{n}}}=\frac{1}{\rho^{\frac{r(p-k)}{p}}}.$$ 
With the change of variables $\lambda_i=\rho^{-\alpha_i}$ $\forall i=1,\ldots,l$, we have $\forall k=1,\ldots,l$
$$d_H^2 \!\dotgeq\!  2\rho^{1-\frac{r}{p}} \rho^{-\frac{1}{k} \sum\limits_{i=1}^k \alpha_i} \rho^{-\frac{r(p-k)}{p}}\!= 2\rho^{-\frac{1}{k}\big(\sum\limits_{i=1}^k \alpha_i +r -k \big)}\!\!=2 \rho^{\delta_k(\boldsymbol\alpha)}$$
where $\boldsymbol\alpha=(\alpha_1,\ldots,\alpha_l)$ and $\delta_k(\boldsymbol\alpha)=-\frac{1}{k}\left(\sum\limits_{i=1}^k \alpha_i +r -k \right)$.\\
Since $2 \norm{W}^2 \sim 2 \chi^2(2mp)$, we have
{\allowdisplaybreaks
\begin{align*}
&P_e(H) \leq \mathbb{P}\left\{ \norm{W}^2 > \frac{\rho^{\delta_k(\ba)}}{2}\right\}\\
&=\sum_{j=0}^{mp-1}e^{-\frac{\rho^{\delta_k(\ba)}}{4}} \left(\frac{\rho^{\delta_k(\ba)}}{4}\right)^j\frac{1}{j!}=\Phi(\delta_k(\ba,r)).
\end{align*}
}
By averaging with respect to the distribution $p(\ba)$, we get
\begin{equation*} 
P_e \leq \int_{\mathcal{A}} p(\boldsymbol\alpha) \Phi(\delta_k(\ba,r))d\boldsymbol\alpha \leq  \int_{\mathcal{A}} p'(\boldsymbol\alpha) \Phi(\delta_k(\ba,r))d\boldsymbol\alpha
\end{equation*}
where $\mathcal{A}=\{\ba: \alpha_1 \leq \cdots \leq \alpha_l\}$, and
$$p'(\boldsymbol\alpha)=K(\log \rho)^l e^{-\sum\limits_{i=1}^l \rho^{-\alpha_i}}\ \rho^{-\sum\limits_{i=1}^l \alpha_i N_i},$$
where $N_i=2(\Delta+2l-2i+1)$. Note that $p'(\ba)$ and $\Phi(\delta_k(\ba,r))$ have the same form as in (\ref{p_prime}) and (\ref{Phi}). From Lemma \ref{laplace_lemma} we find $d(r) \geq \inf_{\ba \in \mathcal{A}_0}  2\sum_{i=1}^l \alpha_i (\Delta+2l-2i+1)$,
which by Lemma \ref{inf_lemma} is the piecewise linear function connecting the points 
$(r,[(n-2r)(m-r)]^+)$ for $r \in \Z$.
\end{IEEEproof}


\appendix

\subsection{Proof of Lemma \ref{inf_lemma}} \label{proof_inf_lemma}
Let $\bar{d}(s)=(-q-l+2\floor{s}+1)s+ql-\floor{s}(\floor{s}+1)$. Without loss of generality, we can suppose that $k-1 \leq s < k$ for some $k \in \N$, i.e. $k-1=\floor{s}$, $k=\floor{s}+1$.\\
First, we show that $\forall \ba \in \mathcal{A}_0$, we have $f(\ba) \geq \bar{d}(s)$. In fact
{\allowdisplaybreaks
\begin{align*}
&f(\ba)=\left(q-l-1\right)\sum\limits_{i=1}^l \alpha_i +2\sum\limits_{i=1}^l (l-i+1)\alpha_i  \\
& \geq \left(q-l-1\right)(l-s)+2\sum\limits_{i=k}^l \sum_{j=1}^i\alpha_i \\
&\geq \left(q-l-1\right)(l-s)+2\sum\limits_{i=k}^l (i-s)\\
&=\left(q-l-1\right)(l-s)+l(l+1)-(k-1)k-2(l-k+1)s\\
&=\bar{d}(s).
\end{align*} 
}
Next, we show that $\exists \ba^*$ such that $f(\ba^*)=\bar{d}(s)$.\\ 
Let $\alpha_1^*=\ldots=\alpha_{k-1}^*=0$, $\alpha_k^*=k-s$, $\alpha_{k+1}^*=\ldots=\alpha_l^*=1$. Then
{\allowdisplaybreaks
\begin{align*}
&f(\ba^*)=\sum_{i=1}^l \left(q+l+1\right)\alpha_i-2\sum_{i=1}^l i \alpha_i \\
&=\left(q+l+1\right)(l-s)-2k(k-s) -l(l+1)+k(k+1)\\
&=\bar{d}(s) \tag*{\IEEEQED}
\end{align*}
}

\subsection{Proof of Lemma \ref{laplace_lemma}} \label{proof_laplace_lemma} 
The proof closely follows \cite{EKPKLold}, which is a preliminary version of \cite{EKPKL}. Note that $\Phi(\rho^{\delta_k(\ba,s)}) \leq 1$ since it is a probability. Given $\varepsilon>0$, we can bound the integral (\ref{integral_A}) as follows
\begin{equation} \label{integral_2}
P_e \leq \int_{\bar{\mathcal{A}}} p'(\boldsymbol\alpha) \Phi(\rho^{\delta_k(\ba,s)})d\boldsymbol\alpha + \sum_{j=1}^l \int_{\mathcal{A}_j} p'(\boldsymbol\alpha) \Phi(\rho^{\delta_k(\ba,s)})d\boldsymbol\alpha,
\end{equation}
where $\bar{\mathcal{A}}=\{\ba \in \mathcal{A}\;:\;\alpha_i \geq -\varepsilon \;\;\forall i=1,\ldots,l\}$ and $\mathcal{A}_j=\{\ba \in \mathcal{A}\;:\;\alpha_j < - \varepsilon\}$. 
Note that
{\allowdisplaybreaks
\begin{align*}
&\int_{\mathcal{A}_j} p'(\boldsymbol\alpha) \Phi(\rho^{\delta_k(\ba,s)})d\boldsymbol\alpha \leq \int_{\mathcal{A}_j} p'(\boldsymbol\alpha)d\boldsymbol\alpha \\
&\leq \left(\prod_{i \neq j} \int_{-\infty}^{\infty} e^{-\rho^{-\alpha_i}} \rho^{-\alpha_i N_i} d\alpha_i\right)\int_{-\infty}^{\varepsilon} e^{-\rho^{-\alpha_j}}\rho^{-\alpha_j N_j} d\alpha_j\\
&=\left(\prod_{i \neq j} \int_{0}^{\infty} \frac{e^{-\lambda_i} \lambda_i^{N_i-1}}{\log \rho}\right)\int_{\rho^{\varepsilon}}^{\infty} \frac{\lambda_j e^{-\lambda_j}}{\log \rho} d\lambda_j \\
&\doteq \rho^{0} \int_{\rho^{\varepsilon}}^{\infty} \frac{\lambda_j e^{-\lambda_j}}{\log \rho} d\lambda_j 
\end{align*}
}
which vanishes exponentially fast as a function of $\rho$. For the first term in (\ref{integral_2}), we have
{\allowdisplaybreaks
\begin{multline*}
\int_{\bar{\mathcal{A}}} p'(\boldsymbol\alpha) \Phi(\rho^{\delta_k(\ba,s)})d\boldsymbol\alpha \leq \int\limits_{\substack{\ba > -\varepsilon \\ \boldsymbol\delta(\alpha,s)< \varepsilon}} p'(\boldsymbol\alpha) \Phi(\rho^{\delta_k(\ba,s)}) d\ba \\
+\sum_{j=1}^n \int\limits_{\substack{\ba > -\epsilon,\\ \delta_j(\ba,s)\geq \varepsilon}}p'(\boldsymbol\alpha) \Phi(\rho^{\delta_k(\ba,s)}) d\ba,
\end{multline*}
}%
where the notation $\ba > -\epsilon$ means $\alpha_i >-\epsilon \;\;\forall i=1,\ldots,l$. We have
{\allowdisplaybreaks
\begin{align}
&\int\limits_{\substack{\ba > -\epsilon,\\ \delta_j(\ba,s)\geq \varepsilon}}p'(\boldsymbol\alpha) \Phi(\rho^{\delta_k(\ba,s)}) d\ba \label{d}\\
&\leq \int\limits_{\substack{\ba > -\epsilon,\\ \delta_j(\ba,s)\geq \varepsilon}} e^{-\frac{\rho^{\delta_j(\ba,s)}}{4}} \sum_{t=0}^{2mn-1}\left(\frac{\rho^{\delta_j(\ba,s)}}{4}\right)^t\frac{1}{t!} \prod_{i=1}^l \rho^{-\alpha_i N_i} d\ba \notag \\
&\leq \left(\prod_{i=j+1}^{l} \int\limits_{\alpha_i >-\varepsilon} \rho^{-\alpha_i N_i} d\alpha_i\right) \notag \\
&\cdot\!\! \int\limits_{\substack{\alpha_1,\ldots,\alpha_j >- \varepsilon\\ \delta_j(\ba,s) \geq \varepsilon}}\!\!e^{-\frac{\rho^{\delta_j(\ba,s)}}{4}}\sum_{t=0}^{2mn-1}\!\!\left(\frac{\rho^{\delta_j(\ba,s)}}{4}\right)^t \frac{1}{t!} \rho^{-\sum\limits_{i=1}^j N_i}d\alpha_1 \ldots d\alpha_j \notag
\end{align}
}
since $\delta_j(\ba,s)$ is independent of $\alpha_i$ for $i >j$. As $\delta_j(\ba,s) \geq \varepsilon$, $\alpha_i>-\varepsilon $ implies $\alpha_i \leq -j\varepsilon -s +j$, the second integral is over a bounded region and tends to zero exponentially fast as a function of $\rho$, while the first integral has a finite SNR exponent. Thus, (\ref{d}) tends to zero exponentially fast. \\
Finally, the SNR exponent of (\ref{integral_A}) is determined by the behavior of 
{\allowdisplaybreaks 
\begin{align*}
&\int\limits_{\substack{\ba > -\varepsilon \\ \boldsymbol\delta(\ba,s)< \varepsilon}} p'(\boldsymbol\alpha) \Phi(\rho^{\delta_k(\ba,s)}) d\ba \leq \int\limits_{\substack{\ba > -\varepsilon \\ \boldsymbol\delta(\ba,s)< \varepsilon}} p'(\boldsymbol\alpha) d\ba \\
&\leq \int\limits_{\substack{\ba > -\varepsilon \\ \boldsymbol\delta(\ba,s)< \varepsilon}} \rho^{-\sum\limits_{i=1}^n N_i \alpha_i} d\ba
\end{align*}
}
The conclusion follows by using the Laplace principle, and taking $\epsilon \to 0$. Note that 
{\allowdisplaybreaks
\begin{align}  
&\mathcal{A}_0=\left\{\ba \in \mathcal{A}:\;\alpha_j\geq 0,\;\sum_{i=1}^j (1-\alpha_i) \leq s \; \forall j=1,\ldots,l\right\} \notag\\
&=\{\ba: \alpha_j \geq 0, \; \delta_j(\ba,s) \leq 0 \;\; \forall j=1,\ldots,l\}. \tag*{\IEEEQED}
\end{align}
}

\begin{footnotesize}
\bibliographystyle{IEEEtran}

\bibliography{STC}

\end{footnotesize}

\end{document}